\documentclass[aps,prb,amssymb,showpacs,twocolumn]{revtex4-2}

\usepackage{graphics,graphicx,amsmath}
\usepackage{tabularx,float}
\usepackage{epsfig}
\usepackage{subfigure}
\usepackage{bm}
\usepackage{wasysym}
\usepackage{bbm}
\usepackage[T1]{fontenc}
\usepackage[french,english]{babel}

\usepackage{mathrsfs} 

\usepackage{color}
\usepackage{verbatim}
\graphicspath{{fig/}}

\begin{document}

\date{\today}

\title{Role of the Berry curvature on BCS-type superconductivity in two-dimensional materials}

\author{Florian Simon}
\affiliation{Laboratoire de Physique des Solides, Universit\'e Paris Saclay,
CNRS UMR 8502, F-91405 Orsay Cedex, France}
\author{Louis Pagot}
\affiliation{Laboratoire de Physique des Solides, Universit\'e Paris Saclay,
CNRS UMR 8502, F-91405 Orsay Cedex, France}
\affiliation{D\'epartement de Physique, \'Ecole Polytechnique, Institut Polytechnique de Paris, Palaiseau, France.}
\author{Marc Gabay}
\affiliation{Laboratoire de Physique des Solides, Universit\'e Paris Saclay,
CNRS UMR 8502, F-91405 Orsay Cedex, France}
\author{Mark O. Goerbig}
\affiliation{Laboratoire de Physique des Solides, Universit\'e Paris Saclay,
CNRS UMR 8502, F-91405 Orsay Cedex, France} 

\begin{abstract}
We theoretically investigate how the Berry curvature, which arises in multi-band structures when the electrons can be described by an effective single-band Hamiltonian, affects the superconducting properties of two-dimensional electronic systems. Generically the Berry curvature is coupled to electric fields beyond those created by the periodic crystal potential. A potential source of such electric fields, which vary slowly on the lattice scale, is the mutual interaction between the electrons. We show that the Berry curvature provides additional terms in the Hamiltonian describing interacting electrons within a single band. When these terms are taken into account in the framework of the usual BCS weak-coupling treatment of a generic attractive interaction that allows for the formation of Cooper pairs, the coupling constant is modified. In pure singlet and triplet superconductors, we find that the Berry curvature generally lowers the coupling constant and thus the superconducting gap and the critical temperature as a function of doping. From an experimental point of view, a measured deviation from the expected BCS critical temperature upon doping, \textit{e.g.} in doped  two-dimensional transition-metal dichalcogenides, may unveil the strength of the Berry curvature. 

\end{abstract}
\maketitle

\section{Introduction}

A large variety of superconducting materials can be theoretically understood within the standard BCS theory proposed by Bardeen, Cooper and Schrieffer \cite{BCSsupra57,tinkham2004introduction}. Within this framework, the metallic electrons of a single, partially filled band are considered to be bound into (Cooper) pairs by a weak attraction, while other bands are discarded based on the premise that they are much more remote in energy than the typical energy scale set by the attractive interaction. Indeed, the attractive interaction between electrons is, within the standard BCS theory, mediated by phonons via the electron phonon coupling. Within the weak-coupling limit, the typical energy scale for superconductivity is then a fraction of the Debye temperature $k_BT_D$ that is itself in the $10...100$ meV range, while the Fermi energy and the typical band gaps are on the order of $\sim 1$ eV \cite{tinkham2004introduction}. In spite of its great success, BCS theory is not capable of explaining all occurrences of superconductivity and finds severe limitations \textit{e.g.} in the case of strongly correlated materials, such as heavy-fermion superconductivity \cite{HF1,reviewheavyfermion} or high-$T_c$ superconductivity \cite{reviewhighTc}, where even the origin of the attractive interaction is still debated. 

While the above-mentioned energy-scale consideration has remained unchallenged for a long time, the advent of topological band theory \cite{TBT,cayssolfuchs} and its success in the theoretical description of a plethora of materials \cite{alltopoallmate}, such as topological insulators \cite{TI1,TI2}, topological superconductors \cite{bernevigBook,TSC}, Weyl and Dirac semimetals \cite{WSM}, has shown that the coupling between energy bands is not only governed by energy scales but by more subtle geometric quantities, such as the Berry curvature or the quantum metric. Several recent papers have investigated the role of the latter, namely in the presence of flat bands in which the quantum metric can be the dominant contribution to the superfluid weight \cite{peotta_superfluidity_2015,rossi2021quantum,tormaberneivig,tian2021evidence}. The Berry curvature has been theoretically shown to play a relevant role in a two-body problem that is closely related to the Cooper pair, namely in the physics of excitons. For example, in two-dimensional (2D) semiconducting transition-metal dichalcogenides (TMDC) \cite{TMDC}, excitons -- bound electron-hole pairs -- are formed in the vicinity of the $K$ and $K'$ points of the first Brillouin zone, where the Berry curvature reaches its maximal value \cite{Fuchs2010}. Experimentally, a first hint to the relevance of band-geometric effects came from the failure of the effective hydrogen model, which had been extremely successful before in the theoretical understanding of the measured exciton spectra \cite{exciton1,exciton2}. It was later shown that the Berry curvature affects the exciton spectra, contrary to the one-particle case, because it couples to the electric field that is generated by the attractive interaction between the electron and the hole forming the bound exciton state \cite{BerryExc1,BerryExc2,Trushin_2017,Hichri_2019}. This is a consequence of the intrinsic Dirac character of the low-energy charge carriers in these materials, which are commonly described in terms of a 2D massive Dirac equation \cite{xiao,GoerbigEPL}. Excitons in 2D TMDC and potentially other bound pairs inherit then this Dirac character \cite{Trushin_2016}.

Based on the above-mentioned exciton example, it is therefore natural to consider that the Berry curvature might also affect the formation of the Cooper pair due to the mutual interaction between the two electrons. This is the main motivation of the present theoretical study, where we show that the effective electron-electron interaction is generically weakened when one includes energy terms in the Hamiltonian that take into account the effect of the Berry curvature. We consider conventional BCS-type superconductivity in 2D materials, such as the above-mentioned 2D semiconducting TMDC for a moderate doping range. We emphasize that we do not investigate topological superconductivity \cite{TSC} that arises when one considers the quasiparticle bands, the mutual coupling of which is at the origin of the emergent topological properties. Here, we rather treat the role of the Berry curvature of the normal state, which affects the formation of Cooper pairs in conventional BCS theory. Within topological band theory, the related wave-vector ($\vec{k}$) dependent Berry connection $\mathcal{A}_n(\vec{k})$ modifies the electrons' positions $\vec{r}$ when the latter are projected by the projectors $P_n$ to the $n$-th band, $\vec{r}\rightarrow P_n \vec{r} P_n=\vec{r}+\mathcal{A}_n(\vec{k})$. This yields a dipole that interacts with the electric field, and this dipolar structure, which the Cooper pair inherits, is at the origin of the weakened Cooper pairing. More precisely, the projection yields two extra terms which affect the electron-electron interaction to the one-body Hamiltonian. One of them is reminiscent of the spin-orbit coupling if one interprets the Berry curvature in terms of a spin, and the second one corresponds to the Darwin term, which arises within a Dirac-fermion treatment of the two bands in the vicinity of the direct gap \cite{FW}. We show that the latter is responsible for a reduced effective BCS coupling constant that results in a smaller superconducting BCS gap, while the former spin-orbit-type term does not play a role in $s$-wave nor other types of pure singlet or triplet pairing. 

The paper is organized as follows. In Sec. \ref{sec:1body}, we briefly revisit, along the lines exposed in Ref. \cite{Hichri_2019}, the emergence of corrective terms to the one-body Hamiltonian of a charge projected to a single band. We present two complementary approaches: one based on a generalized version of the Peierls substitution in Sec. \ref{ssec:Peierls} and one based on a treatment within the continuum two-band model of massive Dirac fermions in the vicinity of the direct gap, where the role of the Berry curvature is most prominent. This treatment is the basis of the two-body problem, which we present in Sec. \ref{sec:2body}. After some general considerations (Sec. \ref{ssec:gen}), Sec. \ref{ssec:Cooper} shows how the Cooper pair and its binding energy are modified by the extra terms, while Sec. \ref{sec:BCS} presents the BCS theory of conventional $s$-wave-type superconductivity in the presence of the corrective terms due to the Berry curvature. In the calculations, we consider a Fermi level that is extremely close to the conduction-band bottom, and we discuss then the role of stronger doping on Cooper pairing and BCS superconductivity in Sec. \ref{sec:doping}. In Sec. \ref{sec:beyond}, we briefly discuss how our theoretical picture of superconductivity in the presence of non-zero Berry curvature evolves in other pairing symmetries, be they singlet or triplet. The last section (Sec. \ref{sec:exp}) is devoted to possible experimental implications of our theoretical studies. There, we compare the superconducting gap and the critical temperature in the absence and the presence of the weakened interaction due to the Berry curvature.

\section{One-body Hamiltonian: corrective terms due to the Berry curvature}
\label{sec:1body}

Before discussing the role of possible geometric terms on the superconducting properties of a 2D material, let us briefly revisit the emergence of these terms within a one-particle description. More precisely, we consider a band structure with $N$ bands described by the Bloch Hamiltonian. The Berry curvature may be viewed as the action of virtual interband transitions of electrons that are otherwise restricted to a single band, while there are no true (quantum) transitions in the adiabatic limit. Notice that there are no geometric terms in the Hamiltonian in the absence of a local electric potential $V(\vec{r})$ different from the periodic one that gives rise to the Bloch bands, and the Hamiltonian is then reduced to the bare band dispersion $E_n(\vec{k})$ of the $n$-th band which the electrons are projected to. 

In the presence of a local potential $V(\vec{r})$ which acts on our single electron, the simple reduction of the Hamiltonian to the band dispersion is no longer valid -- in the following we consider this potential to be generated by the second electron to which the first one is bound in a Cooper pair, but our arguments are not restricted to this case. Indeed, $V(\vec{r})$ couples directly the different bands and needs thus to be taken into account prior to the adiabatic projection to a single band. This yields extra terms to the Hamiltonian that can be discussed within two complementary approaches that we briefly review in this section. The first one is based on a generalized Peierls substitution \cite{gosselin_menas_berard_mohrbach_2006,PhysRevLett.115.166803,Chang2008BerryCO,Gosselin_2008,Trushin_2017,Hichri_2019}. It yields a corrected (quantum) Hamiltonian that reproduces the semi-classical equations of motion. This approach has the advantage of providing a transparent physical interpretation of the role played by the Berry curvature, namely in the formation of a \textit{dipole-like} term that arises due to the projection to a single band. This approach is similar to the magnetic-field case when the electron motion is restricted to a single Landau level \cite{LLdipole0,LLdipole1}, but it does not provide all corrective terms, even at linear order in the Berry curvature. In order to obtain the missing term, which is analogous to the Darwin term in relativistic quantum mechanics, we interpret the Berry curvature in terms of a two-band model, which describes the band structure locally in reciprocal space in terms of a massive Dirac Hamiltonian.

\subsection{Generalized Peierls substitution: emergence of the Berry dipole}
\label{ssec:Peierls}

Let us first recall how to incorporate the magnetic field to describe the dynamics of an electron in the $n$-th band $E_n(\vec{k})$ via the Peierls substitution (in the absence of a Berry curvature). Because the wave vector $\vec{k}=-i\nabla_{\vec{r}}$ is not a gauge-invariant quantity, it needs to be replaced by its gauge-invariant form
\begin{equation}\label{eq:PeierlsK}
\hbar\vec{k}\longrightarrow\vec{\Pi}=\hbar\vec{k}+e\vec{A}(\vec{r}),
\end{equation}
in terms of the vector potential $\vec{A}(\vec{r})$ which yields the magnetic field, $\vec{B}(\vec{r})=\vec{\nabla}_{\vec{r}}\times\vec{A}(\vec{r})$. We consider, here, electrons of charge $-e$ ($e>0$). From a semi-classical point of view, one obtains the equations of motion 
\begin{equation}
    \dot{\vec{r}}_n=\vec{v}_n= \frac{1}{\hbar}\nabla_{\vec{k}}E_n \qquad \text{and}\qquad
    \hbar\dot{\vec{k}}=-e\vec{v}_n \times \vec{B},
\end{equation}
where $\vec{r}_n$ and $\vec{v}_n$ are the average position and velocity, respectively, of the electron in the $n$-th band. One justification of the Peierls substitution is that the Hamiltonian thus obtained, $H(\vec{\Pi})=E_n(\vec{\Pi})$, yields the same equations of motion if one uses the \textit{quantum} Heisenberg equations of motion 
\begin{equation}
    i\hbar \dot{\Pi}_j=[\Pi_j,H(\vec{\Pi})],
\end{equation}
with the help of the commutation relations $[\Pi_x,\Pi_y]=-i\hbar^2/l_B^2$, in terms of the magnetic length $l_B=\sqrt{\hbar/eB}$. Indeed, one then obtains 
\begin{equation}
    \dot{\Pi}_j=-\frac{\hbar}{l_B^2} \epsilon_{jl} \frac{\partial H}{\partial \Pi_l},
\end{equation}
where  $\epsilon_{jl}$ is the antisymmetric Levi-Civita tensor. The quantum Hamiltonian $H(\vec{\Pi})$ yields therefore Heisenberg equations of motion that are the same as the semi-classical ones if we identify the (semi-classical) wave vector $\vec{k}$ with the gauge-invariant quantity $\vec{\Pi}/\hbar$, as it is precisely stipulated by the Peierls substitution. 

The generalized Peierls substitution follows the same spirit when considering a system with a non-zero Berry curvature in the presence of a spatially varying potential $V(\vec{r})$, thus starting from the band energy $H_n=E_n(\vec{k})+V(\vec{r})$. In this case, the semi-classical equations of motion read \cite{Niu,cayssolfuchs}
\begin{eqnarray}\label{eq:semicl}
 \dot{\vec{r}}_n = \vec{v}_n &=& \frac{1}{\hbar}\nabla_{\vec{k}}E_n + \frac{1}{\hbar} \nabla_{\vec{r}}V(\vec{r})\times\vec{\mathcal{B}}_n(\vec{k}) \\ 
\text{and}\qquad \hbar\dot{\vec{k}} &=& -\nabla_{\vec{r}}V-e\vec{v}_n \times \vec{B},
\end{eqnarray}
where $\vec{\mathcal{B}}_n(\vec{k})=\nabla_{\vec{k}}\times \mathcal{A}_n(\vec{k})$ is the Berry curvature of the $n$-th band in terms of its Berry connection $\mathcal{A}_n(\vec{k})$. Similarly to the case discussed above, one can obtain these equations of motion from a \textit{quantum} Hamiltonian \begin{equation}\label{eq:hamKR}
    H(\vec{\Pi},\vec{R})=E_n(\vec{\Pi})+V(\vec{R}),
\end{equation}
where we have replaced not only the wave vector by its gauge-invariant expression (\ref{eq:PeierlsK}) but also the position by its expression projected onto the $n$-th band \cite{Sundaram1999,Niu,cayssolfuchs}
\begin{equation}\label{eq:PeierlsR}
    \vec{r}\longrightarrow\vec{R}=\vec{r}+\vec{\mathcal{A}}_n(\vec{k}),
\end{equation}
which involves the Berry connection $\mathcal{A}_n(\vec{k})$. Similarly to the Peierls substitution (\ref{eq:PeierlsK}), the position $\vec{r}$ on the right-hand-side of this expression should be interpreted as a reciprocal-space derivative $\vec{r}=i\nabla_{\vec{k}}$. The replacement (\ref{eq:PeierlsR}) may be viewed as a \textit{generalized Peierls substitution} \cite{gosselin_menas_berard_mohrbach_2006,PhysRevLett.115.166803,Chang2008BerryCO,Gosselin_2008,Hichri_2019}. The semi-classical equations of motion are then retrieved as the Heisenberg equations of motion not only for $\vec{\Pi}$ but also for $\vec{R}=(X,Y)$ on the basis of the Hamiltonian (\ref{eq:hamKR}) and the induced commutation relations $[X,Y]=i\mathcal{B}_n(\vec{k})$ \cite{Hichri_2019}.

Let us now discard the magnetic field, which we have only discussed in order to remind the reader of the Peierls substitution and to justify its generalized form and expand the Hamiltonian (\ref{eq:hamKR}) to lowest order in the Berry connection. This expansion is legitimate as long as the external potential $V(\vec{r})$ varies slowly on a length scale that is set, in orders of magnitude, by the Berry connection and that can be related to an effective Compton length, as we discuss below. The Hamiltonian then becomes 
\begin{equation}\label{eq:hamPeierls}
    H=E_{n}(\vec{k})+V(\vec{r})+\vec{\mathcal{A}}_n(\vec{k})\cdot\vec{\nabla}_{\vec{r}}V(\vec{r}).
\end{equation}
The last generated term is interesting.  First, it can be interpreted as the energy of an electric dipole $-e\vec{\mathcal{A}}_n(\vec{k})$ in an electric field $\vec{E}(\vec{r}) = \nabla V(\vec{r})/e$. We therefore call this term the \textit{Berry dipole term}.   
Second, this term can be understood as an effective spin-orbit coupling if we use the \textit{symmetric gauge} for the Berry connection
\begin{equation}
    \vec{\mathcal{A}}_n(\vec{k})=\frac{1}{2}\vec{\mathcal{B}}_n(\vec{k})\times\vec{k},
\end{equation}
in which case the corrective term reads
\begin{equation}
    \vec{\mathcal{A}}_n(\vec{k})\cdot\vec{\nabla}_{\vec{r}}V(\vec{r})=\frac{1}{2}\Big(\vec{\mathcal{B}}_n(\vec{k})\times\vec{k}\Big)\cdot\vec{\nabla}_{\vec{r}}V(\vec{r}).
\end{equation}
This expression is interesting for the following reason. The Berry curvature is often viewed as the analogue of a magnetic field in reciprocal space, while the extra term in Eq. (\ref{eq:hamPeierls}) has the same form as the spin-orbit coupling term, which arises when one projects the relativistic Dirac equation onto the electron (or positron) branch \cite{greiner2000}. In this analogy, one would however need to identify the Berry curvature with an emergent spin rather than with a magnetic field.

\subsection{Non-relativistic limit of the Dirac equation}
\label{ssec:Dirac}

\begin{figure}[h!]
    \label{figbandes}
        \centering
        \includegraphics[width=0.4\textwidth]{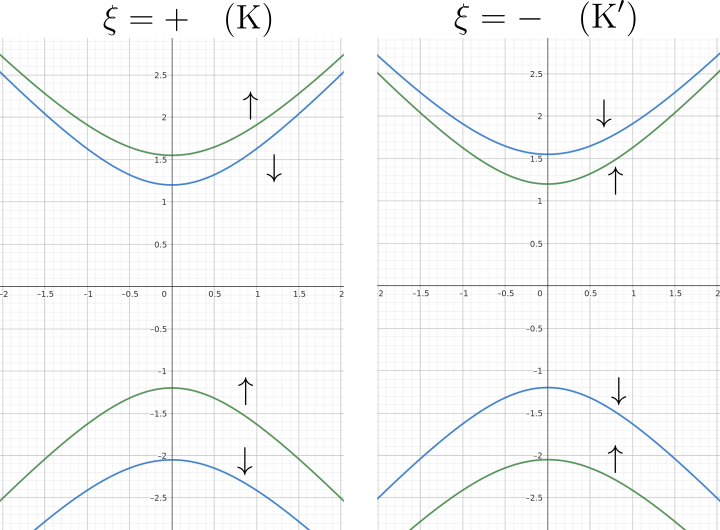}
        \caption{Band structure of massive Dirac fermions, with \textit{a priori} two different gaps for the two values of $\xi\sigma$, as one typically encounters in 2D semiconducting TMDC.}
        \label{fig:01}
\end{figure}

In many situations the role of the Berry curvature in semiconducting materials can be approached in terms of a massive Dirac equation that describes two coupled bands in the vicinity of a reciprocal-space point, where the band gap is smallest and the Berry curvature has a maximum \cite{DiracBerry,Fuchs2010}. In this picture, coupling to other bands is not \textit{per se} excluded, but we consider that it only gives rise to a negligible contribution to the respective Berry curvatures of the two bands. This situation arises, \textit{e.g.}, in 2D semi-conducting TMDC in which two spin-orbit coupled families of band pairs form a direct gap at the $K$ and $K'$ points. In the vicinity of these points, the two bands are described by the generic Dirac Hamiltonian
\begin{equation}\label{eq:hamDir}
    H=
    \begin{pmatrix}
    \Delta_{\xi\sigma} \sigma_0 & \hbar v_D(\xi\sigma k_x-ik_y) \\
    \hbar v_D(\xi\sigma k_x +i k_y) & -\Delta_{\xi\sigma}\sigma_0
    \end{pmatrix} + E_{\xi\sigma}^0 + V(\vec{r}),
\end{equation}
where $\xi$ indicates the valley index ($\xi=+$ for the $K$ valley and $\xi=-$ for the $K'$ valley in the case of 2D TMDC, or generally two time-reversal-symmetry related points $\pm \vec{k}_D$) and $\sigma=\pm$ represents the physical spin. In the presence of spin-orbit coupling and time-reversal symmetry, the band gaps $2\Delta_{\xi\sigma}$ of the two valleys are locked and depend only on the product $\xi\sigma$ of the spin and valley index, and so does the shift in energy $E_{\xi\sigma}^0$, which does not play any topological or dynamical role. In the absence of the external potential $V(\vec{r})$, one obtains the four bands
\begin{equation}
    \epsilon_{\lambda,\xi\sigma}(\vec{k})=E_{\xi\sigma}^0+\lambda \sqrt{\Delta_{\xi\sigma}^2+\big(\hbar v_Dk)^2}, 
\end{equation}
which is depicted in Fig. \ref{fig:01}. The index $\lambda $ refers to the conduction ($\lambda=+$) and the valence ($\lambda=-$) bands. Note that there are only four bands since spin and valley are locked -- they enter into the expressions only as the product label $\xi\sigma$ -- as it is required by time-reversal symmetry. The associated Berry curvatures are given by \cite{DiracBerry,Niu}
\begin{equation}\label{eq:Berry}
    \vec{\mathcal{B}}_{\lambda,\xi\sigma}(\vec{k})=-\frac{\lambda\xi\sigma}{2}\frac{\lambdabar_{\xi\sigma}^2}{\big(1+\lambdabar_{\xi\sigma}^2k^2\big)^{3/2}}\vec{e}_z\qquad\lambdabar_{\xi\sigma}=\frac{\hbar v_D}{\Delta_{\xi\sigma}},
\end{equation}
where $\vec{e}_z$ denotes the unit vector in the $z$-direction. The last expression $\lambdabar_{\xi\sigma}$ represent the characteristic length scale, which we have already mentioned in the previous subsection and that yields the order of magnitude for the displacement and thus the dipole as a consequence of projection onto a single band. It is inversely proportional to the band gap $\Delta_{\xi\sigma}$ and constitutes a lower bound for all length scales. It is reminiscent of the Compton length in high-energy physics \cite{Compton,greiner2000}. Indeed, if we rewrite the gap in terms of the band masses $m_{\xi\sigma}$, $\Delta_{\xi\sigma}=m_{\xi\sigma}v_D^2$, one retrieves its more familiar form $\lambdabar_{\xi\sigma} =\hbar/m_{\xi\sigma}v_D$. Physically it represents a limiting length below which the Compton effect transforms erratically photons into electron-positron pairs, so that information encoded in the phase of the light field can no longer be used for spectroscopic means. In condensed-matter physics, the interpretation of this length is similar: processes of characteristic length scales below $\lambdabar_{\xi\sigma}$ inevitably yield interband transitions that drive the system out of the regime of validity of the adiabatic approximation, which provided us with the semi-classical equations of motion (\ref{eq:semicl}).

For transport properties, including superconductivity, the most important electrons are those in the vicinity of the Fermi level, which we consider here to be close to the bottom of the conduction band, \textit{i.e.} we consider a moderately doped semiconductor. We can already anticipate that the Berry curvature may play a role as long as the Fermi wave vector $k_F$ satisfies $\lambdabar_{\xi\sigma} k_F\ll 1$ since it vanishes algebraically for $\lambdabar_{\xi\sigma}\rightarrow \infty$ [see Eq. (\ref{eq:Berry})]. We therefore project the Hamiltonian (\ref{eq:hamDir}) onto the conduction-band bottom, $0<\delta E=E-\Delta_{\xi\sigma}-E_{\xi\sigma}^0\ll \Delta_{\xi\sigma}$ (see Fig. \ref{fig:01}), with the help of the Foldy-Wouthuysen transformation to keep track of the electric potential $V(\vec{r})$ \cite{FW}. 
This yield the effective one-band Hamiltonian
\begin{eqnarray}\label{eq:hamPauli}
\nonumber
     H &\simeq& E_{\xi\sigma}^0+\Delta_{\xi\sigma}+\frac{\hbar^2\vec{k}^2}{2m_D}+ V(\vec{r})\\
     &&+\frac{\xi\sigma\lambdabar_{\xi\sigma}^2}{4}\Big(\vec{e}_z\times\vec{k}\Big)\cdot\vec{\nabla}_{\vec{r}}V+\frac{\lambdabar_{\xi\sigma}^2}{8}\vec{\nabla}_{\vec{r}}^2V  ,
\end{eqnarray}
which, apart from the last term, is identical to the one (\ref{eq:hamPeierls}) which we have obtained with the help of the generalized Peierls subsitution if we make use of the expression (\ref{eq:Berry}) for the Berry curvature to lowest order in the wave vector and if we redefine the energy with respect to the band bottom. The last term may also be written in terms of the Berry curvature as 
\begin{equation}
    \frac{\lambdabar_{\xi\sigma}^2}{8}\vec{\nabla}_{\vec{r}}^2V(\vec{r})=\frac{1}{4}\big|\mathcal{B}_{\lambda,\xi\sigma}(0)\big|\vec{\nabla}^2_{\vec{r}}V(\vec{r})
\end{equation}
and corresponds to the Darwin term in high-energy physics. While it does not play any role in the semi-classical equations of motion, it is relevant namely at very short ranges and has been shown to strongly affect \textit{e.g.} the spectra of $s$-state excitions in 2D TMDC \cite{BerryExc1,BerryExc2,Trushin_2017,Hichri_2019}. This is best seen in the case of the 2D Coulomb potential in which case $\nabla^2_{\vec{r}}V=e^2\delta(\vec{r})/\epsilon$, \textit{i.e.} it is relevant for pair wave functions with a non-zero amplitude at the origin ($s$-wave states) such as the BCS wave functions, which we discuss below.  

\section{Two-body problem: General case and Cooper pair}
\label{sec:2body}

With the Cooper-pair problem in mind, we now consider how the extra terms discussed within the one-particle picture presented in the preceding section evolves in the case of two electrons at the bottom of the conduction band $\lambda=+$ at the same energy. This choice to consider a Fermi level slightly above the bottom of the conduction band is perfectly arbitrary, but the results obtained in the following sections remain valid for Cooper pairs formed from holes in the valence band. We consider again the spin to be locked to the valley index so that there is only one effective label $\xi\sigma$, which we represent by the valley index ($\xi_1$ for the first electron and $\xi_2$ for the second one) to simplify the notations. Furthermore we consider a two-body potential $V$ that depends only on the relative position of the two electrons $\vec{r}_1-\vec{r}_2$, such as it is the case for the BCS potential. 

\subsection{General case}\label{ssec:gen}

Because the two-body interaction potential only depends on the relative distance   $\vec{\rho}=\vec{r}_1-\vec{r}_2$ between the electrons, we introduce relative and center-of-mass (CoM) coordinates. Since both electrons have the same mass, we have
\begin{align}
    \text{Relative:}\qquad&\vec{\rho}=\vec{r}_1-\vec{r}_2\hspace{2cm}\vec{k}=\frac{\vec{k}_1-\vec{k}_2}{2} \\
    \text{CoM:}\qquad&\vec{R}=\frac{\vec{r}_1+\vec{r}_2}{2}\hspace{2cm}\vec{K}=\vec{k}_1+\vec{k}_2,
\end{align}
Separation of the CoM and relative coordinates yields the Hamiltonian 
\begin{widetext}
\begin{eqnarray}
\nonumber
    H_{2e^-}&=&2\Delta_b+\frac{\hbar^2\vec{K}^2}{4m_D}+\frac{\hbar^2\vec{k}^2}{m_D}+V(\vec{\rho})+\frac{1}{4}\bigg(\vec{\Lambda}^{\xi_1,\xi_2}_{+}\big(\vec{K},\vec{k}\big)\times\vec{K}\bigg)\cdot\vec{\nabla}V(\vec{\rho})+\frac{1}{2}\bigg(\vec{\Lambda}^{\xi_1,\xi_2}_{-}\big(\vec{K},\vec{k}\big)\times\vec{k}\bigg)\cdot\vec{\nabla}V(\vec{\rho}) \\
    &&+\frac{1}{2}\big|\mathcal{B}(0)\big|\vec{\nabla}^2V(\vec{\rho}) 
    \qquad\text{with}\quad\vec{\Lambda}^{\xi_1,\xi_2}_\pm\big(\vec{K},\vec{k}\big)=\vec{\mathcal{B}}_{+,\xi_1}\bigg(\frac{1}{2}\vec{K}+\vec{k}\bigg)\pm\vec{\mathcal{B}}_{+,\xi_2}\bigg(\frac{1}{2}\vec{K}-\vec{k}\bigg) \label{eq:ham2body}
\end{eqnarray}
\end{widetext}
within the parabolic approximation, and where we have made use of the Dirac mass $m_D=\Delta_{\xi\sigma}/v_D^2$. Since we no longer consider $k$-space gradients, we omit the index $\vec{r}$ at the gradient $\nabla_{\vec{r}}=\nabla$ from now on. It is interesting to notice that, when moving to CoM/relative coordinates, the Berry dipole term splits into two dipoles acting on the electron pair. One is associated with its \emph{center-of-mass motion} and the \emph{sum} of the two Berry curvatures and the other is associated with its \emph{relative motion} and the \emph{difference} of the two Berry curvatures. To gain further insight into the physical meaning of these two terms, we can calculate the Heisenberg equations of motion
\begin{align}
    &\overset{.}{\vec{K}} = \vec{0} \qquad
    \hspace{9mm}
    \overset{.}{\vec{R}} = \frac{\hbar\vec{K}}{2m_D}+\frac{1}{4\hbar}\vec{\nabla}V(\rho)\times\vec{\Lambda}^{\xi_1,\xi_2}_+(\vec{K},\vec{k}) \\
    & \overset{.}{\vec{k}} = -\frac{1}{\hbar}\vec{\nabla}H_{2e^-}\qquad \overset{.}{\vec{\rho}}=2\frac{\hbar\vec{k}}{m_D}+\frac{1}{2\hbar}\vec{\nabla}V(\vec{\rho})\times\vec{\Lambda}^{\xi_1,\xi_2}_-(\vec{K},\vec{k})
\end{align}
The CoM momentum is a conserved quantity, owing to the fact that $H_{2e^-}$ does not depend on $\vec{R}$. We also see that the two dipoles induce two Karplus-Luttinger-type velocities: $\Lambda_+$, which is associated to the CoM dipole, generates a drift velocity of the CoM coordinate, and $\Lambda_-$, which is associated to the relative dipole, yields another drift velocity of the relative coordinate of the Cooper pair.

Before discussing the special case of the Cooper pair, we may already discuss here the relative role of the two quantities $\vec{\Lambda}_+$ and $\vec{\Lambda}_-$ as a function of the two different valleys, \textit{i.e.} in the case of \textit{intra-valley} pairing as compared to \textit{inter-valley} pairing. Indeed, they determine the dipolar moments 
\begin{equation}
\vec{d}_{\pm}=-e(\vec{\Lambda}_{\pm}\times \vec{q})/2, 
\end{equation}
where $\vec{q}=\vec{K}$ for the CoM dipole (sign $+$) and $\vec{q}=\vec{k}$ for the relative dipole (sign $-$). In the case of intra-valley pairing ($\xi_1=\xi_2$), which corresponds to triplet superconductivity as a consequence of the spin-valley locking, the relative dipole $\vec{d}_-$ is negligible to lowest order in the wave vectors while the CoM dipole is on the order of $\vec{d}_+\sim -e \mathcal{B}_{+,\xi_1}(0)\times \vec{K}$. Their roles are inverted in the case of singlet-type inter-valley pairing, in which case $\vec{d}_+\simeq 0$ while $\vec{d}_-\sim -e \mathcal{B}_{+,\xi_1}(0)\times \vec{k}$.

\subsection{Revisiting the Cooper problem}\label{ssec:Cooper}

We are now in a position to study the effect of the Berry curvature on a Cooper pair, the building block of superconductors. To do so, we revisit the Cooper problem following the lines of Ref.  \cite{leonn.cooper1956} and standard textbooks \cite{tinkham2004introduction}. The Hamiltonian we consider here is $H_c= H_{2e^-}(\vec{K}=\vec{0})$, \textit{i.e.} our two-body Hamiltonian (\ref{eq:ham2body}) in the rest frame,
\begin{align}
    H_{c}=2\epsilon_+(\vec{k})+V(\vec{\rho})&+\frac{1}{2}\bigg(\vec{\Lambda}^{\xi_1,\xi_2}_{-}\big(\vec{0},\vec{k}\big)\times\vec{k}\bigg)\cdot\vec{\nabla}V(\vec{\rho})\nonumber \\
    &+\frac{1}{2}\big|\mathcal{B}(0)\big|\vec{\nabla}^2V(\vec{\rho}),
\end{align}
where $\Lambda_-$ can be rewritten as
\begin{equation}
    \vec{\Lambda}_-^{\xi_1,\xi_2}(\vec{0},\vec{k})=-(\xi_1-\xi_2)\frac{\lambdabar_{\xi\sigma}^2}{2\big(1+\lambdabar_{\xi\sigma}^2k^2\big)^{3/2}}\vec{e}_z=\vec{\Lambda}_-^{\xi_1,\xi_2}(\vec{k}).
\end{equation}
As mentioned above, one notices that, for the Berry dipole term to be non-zero, the two electrons of the Cooper pair need to be taken in different valleys and thus with opposite spin, as it is usual for $s$-wave singlet superconductivity. In contrast to this, we have $\Lambda_+^{\xi_1,\xi_2}(0,\vec{k})\propto(\xi_1+\xi_2)$ \textit{i.e.} one needs electrons in the same valley, but even then, the intra-valley CoM dipolar term in the Hamiltonian vanishes unless $\vec{K}\neq 0$. We therefore consider henceforth only the relative dipolar term and the case of inter-valley pairing. 

Let us now take a closer look  at the wave function of the Cooper pair $\psi(\vec{\rho})$, which is a solution of $H_c\psi(\vec{\rho})=E\psi(\vec{\rho})$. We then decompose $\psi$ and $V$ in a Fourier series
\begin{align}
    &\psi(\vec{\rho})=\sum_{\vec{k}}g_{\vec{k}}e^{i\vec{k}\cdot\vec{\rho}},\\
    &V(\vec{\rho})=\sum_{\vec{k}\vec{k'}}V_{\vec{k}\vec{k'}}e^{i(\vec{k}-\vec{k'})\cdot\vec{\rho}}\quad.
\end{align}
Following the steps of Ref. \cite{tinkham2004introduction} we find the self-consistent equation
\begin{equation}
    \big[E-2\epsilon_+(\vec{k})\big]g_{\vec{k}}=\sum_{\vec{k}'}V^{\text{eff}}_{\vec{k}\vec{k'}}g_{\vec{k'}}
\end{equation}
for the coefficients $g_{\vec{k}}$, in terms of the \textit{effective interaction}
\begin{equation}
\label{inteff}
    V^{\text{eff}}_{\vec{k}\vec{k}'}=\Bigg[1+\frac{i}{2}\bigg(\vec{\Lambda}^{\xi_1,\xi_2}_{-}(\vec{k})\times\vec{k}\bigg)\cdot\vec{k}'-\frac{1}{2}\big|\mathcal{B}(0)\big|\big(\vec{k}-\vec{k}'\big)^2\Bigg]V_{\vec{k}\vec{k}'}.
\end{equation}
This equation is one of the main results of our paper. Qualitatively, we see that the two terms appear with opposite signs. The second term stems from the Berry dipole term in Hamiltonian (\ref{eq:ham2body}) and may increase or decrease the interaction potential and thus the strength of the Cooper pairing depending on the sign of $\vec{\Lambda}_-$. As for the last (Darwin) term, it is negative irrespective of the valley index, meaning that it tends to weaken the electron-electron interaction and thus the superconducting phase. On a more practical level, the above expressions tell us that the calculations for the energy of the Cooper pair in the presence of a Berry curvature are the same as in the conventional pairing case \cite{tinkham2004introduction}, but in terms of the effective interaction (\ref{inteff}). 

In a second step we need to solve the self-consistency equation
\begin{equation}
    \label{Cooper1}
    \sum_{\vec{k}}\frac{\langle V^{\text{eff}}_{\vec{k}\vec{k'}}\rangle}{E-2\epsilon_+(\vec{k})}=1,
\end{equation}
where we have defined the average
\begin{equation}
    \langle\mathcal{O}(\vec{k'})\rangle=\frac{\sum_{\vec{k'}}\mathcal{O}(\vec{k'})g_{\vec{k'}}}{\sum_{\vec{k'}}g_{\vec{k'}}}
\end{equation}
with respect to the weighting coefficients $g_{\vec{k}}$. The term $\langle V^{\text{eff}}_{\vec{k}\vec{k'}}\rangle$ may be rewritten as
\begin{widetext}
\begin{equation}
    \langle V^{\text{eff}}_{\vec{k}\vec{k'}}\rangle=\bigg(1-\frac{1}{2}\big|\mathcal{B}(0)\big|\vec{k}^2\bigg)\langle V_{\vec{k}\vec{k'}}\rangle+\bigg(\big|\mathcal{B}(0)\big|\vec{k}+\frac{i}{2}\vec{\Lambda}_-^{\xi_1,\xi_2}(\vec{k})\times\vec{k}\bigg)\cdot\langle\vec{k'}V_{\vec{k}\vec{k'}}\rangle-\frac{1}{2}\big|\mathcal{B}(0)\big|\langle\vec{k'}^2V_{\vec{k}\vec{k'}}\rangle.
\end{equation}
\end{widetext}
To illustrate the role of the additional terms due to the Berry curvature, let us consider the BCS potential, defined as
\begin{equation}
    V_{\vec{k}\vec{k'}}=
    \begin{cases}
    &-V<0\quad\text{if }\epsilon_F\leq\epsilon_+(\vec{k}),\epsilon_+(\vec{k'})\leq\epsilon_F+\hbar\omega_D \\
    &0\quad\text{otherwise},
    \end{cases}
\end{equation}
where $\epsilon_F$ is the Fermi energy and $\hbar\omega_D$ the Debye energy. We can compactly rewrite it as 
\begin{equation}
    V_{\vec{k}\vec{k'}}=-V\mathbbm{1}_{\mathcal{D}}(\vec{k})\mathbbm{1}_{\mathcal{D}}(\vec{k'})
    \label{VBCS}
\end{equation}
where $\mathbbm{1}_{\mathcal{D}}$ is the indicator function of the set 
\begin{equation}
    \label{Dset}
    \mathcal{D}=\Big\{\vec{k}\in\mathbb{R}^2\Big|\epsilon_F\leq\epsilon_+(\vec{k})\leq\epsilon_F+\hbar\omega_D\Big\}.
\end{equation}
With this in mind, we write
\begin{equation}
\label{sumPeierls}
    \langle\vec{k'}V_{\vec{k}\vec{k'}}\rangle\propto\sum_{\vec{k'}\in\mathcal{D}}\vec{k'}V_{\vec{k}\vec{k'}}g_{\vec{k'}}
\end{equation}
From Eq. (\ref{VBCS}) we see that $V_{\vec{k};-\vec{k'}}=V_{\vec{k}\vec{k'}}$. Moreover, for BCS superconductivity we have $g_{-\vec{k'}}=g_{\vec{k'}}$ so that $\vec{k'}V_{\vec{k}\vec{k'}}g_{\vec{k'}}$ is an odd function of $\vec{k'}$. Because summing an odd function over the set $\mathcal{D}$ gives zero, we have $\langle\vec{k'}V_{\vec{k}\vec{k'}}\rangle=\vec{0}$ so that \emph{the Berry dipole term does not affect the Cooper pair}, which is then solely affected by the Darwin term. Therefore, if we remember the competition between the dipolar and Darwin terms, this suggests that the effect of the Berry curvature is to weaken the Cooper.

As for $\langle V^{\text{eff}}_{\vec{k}\vec{k'}}\rangle$, we are left with
\begin{equation}
    \langle V^{\text{eff}}_{\vec{k}\vec{k'}}\rangle=\bigg(1-\frac{1}{2}\big|\mathcal{B}(0)\big|\vec{k}^2\bigg)\langle V_{\vec{k}\vec{k'}}\rangle -\frac{1}{2}\langle \big|\mathcal{B}(0)\big|\vec{k'}^2V_{\vec{k}\vec{k'}}\rangle
\end{equation}
Remember that $V^{\text{eff}}_{\vec{k}\vec{k'}}$ is non-zero only for $\vec{k},\vec{k'}\in\mathcal{D}$, and from the definition of $\mathcal{D}$ we rewrite the energy as $\epsilon_+(\vec{k})=\epsilon_F+\eta_{\vec{k}}\hbar\omega_D$ with $\eta_{\vec{k}}\in[0,1]$. Furthermore, suppose that $\hbar\omega_D\ll\epsilon_F-\Delta_{\xi\sigma}$ so that the perturbation does the reach the bottom of the conduction band.
From this and the expression of $\epsilon_+(\vec{k})$ we obtain
\begin{equation}
    \big|\mathcal{B}(0)\big|\vec{k}^2=\frac{\epsilon_F-\Delta_{\xi\sigma}}{\Delta_{\xi\sigma}}+\eta_{\vec{k}}\frac{\hbar\omega_D}{\Delta_{\xi\sigma}}.
\end{equation}
Now, for many 2D materials (including any TMDC), the band gap is in the  1eV range (see e.g. Ref. \cite{doi:10.1021/acs.jpclett.5b01686}) while for most crystals $\hbar\omega_D\sim 0.01$eV \cite{LI2012197}. One therefore obtains a ratio $\frac{\hbar\omega_D}{\Delta_b}\sim 0.01$, so that we may neglect the corresponding term and thus make the approximation
\begin{equation}
\label{approxberry}    \big|\mathcal{B}(0)\big|k^2\simeq\big|\mathcal{B}(0)\big|k_F^2\qquad \big|\mathcal{B}(0)\big|k'^2\simeq\big|\mathcal{B}(0)\big|k_F^2.
\end{equation}
With this and $\langle\vec{k'}V_{\vec{k}\vec{k'}}\rangle=\vec{0}$, we finally obtain
\begin{equation}
    \label{approxint}
    \langle V^{\text{eff}}_{\vec{k}\vec{k'}}\rangle=\big(1-\big|\mathcal{B}(0)\big|k_F^2\big)\langle V_{\vec{k}\vec{k'}}\rangle,
\end{equation}
in line with our qualitative argument of a weakening of the electron-electron interaction induced by the Darwin term. With the BCS potential, $\langle V_{\vec{k}\vec{k'}}\rangle=-V$, one finds
\begin{equation}
    \sum_{\vec{k}}\frac{1}{E-2\epsilon_+(\vec{k})}=-\frac{1}{\big(1-\big|\mathcal{B}(0)\big|k_F^2\big)V}.
\end{equation}

As usual, the sum over the wave vector may be replaced by an integral over energy with the help of the density of states $\rho(\epsilon)$ and the BCS coupling constant $\lambda= V\rho(\epsilon_F)$. We finally find the binding energy of the Cooper pair
\begin{equation}\label{eq:binding}
    E_B=\frac{2\hbar\omega_D}{e^{2/\lambda_{\text{eff}}}-1}\quad\text{with}\quad\lambda_{\text{eff}}=\big(1-\big|\mathcal{B}(0)\big|k_F^2\big)\lambda,
\end{equation}
which is the same as the conventional expression 
\begin{equation}
    E_B^{\text{BCS}}=\frac{2\hbar\omega_D}{e^{2/\lambda}-1},
\end{equation}
where we have replaced $\lambda$ by an effective (lower) coupling constant. If we set the Berry curvature to zero or if we set the band gap to be infinity, we recover the usual expression, as expected.

To summarize this subsection, we highlight two aspects. First, the effect of the Berry curvature on the Cooper pair reveals itself through a competition between two terms. On the one hand, the Berry dipole term, with its dipolar/spin-orbit form, induces a drift velocity analogous to the Karplus-Luttinger veloctity on the relative position of the electrons of the Cooper pair. It \emph{could} in principle enhance the electron-electron interaction $V_{\vec{k}\vec{k'}}$. On the other hand, the Darwin term yields a negative contribution and thus weakens the effective interaction. Second, the Berry dipole term's contribution to Cooper pairing turns out to be zero for $s$-wave superconductivity, and thus we are only left with a weakened electron-electron interaction due to the Darwin term. This is clearly seen in the expression of the binding energy (\ref{eq:binding}) Indeed, since the interaction $V$ is lowered, so is the BCS coupling $\lambda$, thereby lowering the binding energy of the Cooper pair. In conclusion, the Berry curvature makes the Cooper pairs less bound and thus more easily breakable, \textit{e.g.} by thermal fluctuations. This means that the critical temperature (and the superconducting gap) are lowered as well, as we show explicitely in the following section, where we discuss the action of the Berry-curvature corrective terms in the BCS many-body approach. 

\section{BCS Hamiltonian in the presence of Berry curvature}\label{sec:BCS}

In the previous section, we found that the calculations in the electron pair problem with Berry curvature were the same as in its absence, but with an effective interaction. We therefore consider, in this part, the BCS Hamiltonian where we replace the interaction $V_{\vec{k}\vec{k'}}$ with the effective one $V^{\text{eff}}_{\vec{k}\vec{k'}}$ which is given in Eq. (\ref{inteff}) and that accounts for the corrective terms due to the Berry curvature.
\begin{equation}
    H=\sum_{\vec{k}\sigma}\xi_{\vec{k}}c^\dagger_{\vec{k}\sigma}c_{\vec{k}\sigma}+\sum_{\vec{k}\vec{k'}}V^{\text{eff}}_{\vec{k}\vec{k'}}c^\dagger_{\vec{k'}\uparrow}c^\dagger_{-\vec{k'}\downarrow}c_{\vec{k}\uparrow}c_{-\vec{k}\downarrow}
\end{equation}
where $\xi_{\vec{k}}=\epsilon_+(\vec{k})-\epsilon_F$, and the bare interaction (in the absence of Berry curvature corrections) is $V_{\vec{k}\vec{k'}}=-V\mathbbm{1}_{\mathcal{D}}(\vec{k})\mathbbm{1}_{\mathcal{D}}(\vec{k'})$ with $\mathcal{D}=\Big\{\vec{k}\in\mathbb{R}^2\Big|\epsilon_F-\hbar\omega_D\leq\epsilon_+(\vec{k})\leq\epsilon_F+\hbar\omega_D\Big\}$. We also keep the same groundstate. Since this Hamiltonian has the same form as the original BCS Hamiltonian, the same calculations hold as long as the interaction is not specified. We thus find the textbook gap equation \cite{tinkham2004introduction}
\begin{equation}
    \Delta_{\vec{k}}=-\frac{1}{2}\sum_{\vec{k'}}V^{\text{eff}}_{\vec{k}\vec{k'}}\frac{\Delta_{\vec{k'}}}{\sqrt{\Delta_{\vec{k'}}^2+\xi^2_{\vec{k'}}}}\tanh\bigg(\frac{\beta}{2}\sqrt{\Delta_{\vec{k'}}^2+\xi^2_{\vec{k'}}}\bigg)
\end{equation}
with $\Delta_{\vec{k}}=-\sum_{\vec{k'}}V^{\text{eff}}_{\vec{k}\vec{k'}}\langle c^\dagger_{\vec{k'}\uparrow}c^\dagger_{-\vec{k'}\downarrow}\rangle$ and $\beta=(k_BT)^{-1}$. In terms of the auxiliary function
\begin{equation}
    \label{kernel}
    f_{\beta,\vec{k}}(\vec{k'})=\frac{\Delta_{\vec{k'}}}{\sqrt{\Delta_{\vec{k'}}^2+\xi^2_{\vec{k'}}}}\tanh\bigg(\frac{\beta}{2}\sqrt{\Delta_{\vec{k'}}^2+\xi^2_{\vec{k'}}}\bigg),
\end{equation}
the self-consistent gap equation reads
\begin{widetext}
\begin{equation}
    \Delta_{\vec{k}}=-\frac{1}{2}\bigg(1-\frac{1}{2}\big|\mathcal{B}(0)\big|\vec{k}^2\bigg)\sum_{\vec{k'}}V_{\vec{k}\vec{k'}}f_{\beta,\vec{k}}(\vec{k'})-\frac{1}{2}\bigg(\frac{i}{2}\vec{\Lambda}^{\xi_1,\xi_2}_-(\vec{k})\times\vec{k}+\big|\mathcal{B}(0)\big|\vec{k}\bigg)\cdot\sum_{\vec{k'}}\vec{k'}V_{\vec{k}\vec{k'}}f_{\beta,\vec{k}}(\vec{k'})-\frac{1}{2}\sum_{\vec{k'}}\frac{1}{2}\big|\mathcal{B}(0)\big|\vec{k'}^2V_{\vec{k}\vec{k'}}f_{\beta,\vec{k}}(\vec{k'})
\end{equation}
\end{widetext}
One can show that if the bare superconducting gap has a definite parity, then $\Delta_{\vec{k}}$ (so defined through the effective interaction) has the same parity. Therefore for BCS superconductivity we have $\Delta_{-\vec{k}}=\Delta_{\vec{k}}$. From equation (\ref{kernel}), it is then clear that $f_{\beta,\vec{k}}(-\vec{k'})=f_{\beta,\vec{k}}(\vec{k'})$. And since $V_{\vec{k};-\vec{k'}}=V_{\vec{k};\vec{k'}}$, the function $\vec{k'}\longrightarrow\vec{k'}V_{\vec{k}\vec{k'}}f_{\beta,\vec{k}}(\vec{k'})$ is an odd function so that
\begin{equation}
    \sum_{\vec{k'}}\vec{k'}V_{\vec{k}\vec{k'}}f_{\beta,\vec{k}}(\vec{k'})=\vec{0},
\end{equation}
and thus the Berry dipole term does again not affect the many-body result, which is consistent with the results obtained in the previous section. We then make the same approximate treatment [see Eqs. (\ref{approxberry}) and (\ref{approxint})] as for the Cooper pair problem and we find
\begin{equation}
    \Delta_{\vec{k}}=-\frac{1}{2}\sum_{\vec{k'}}\big(1-\big|\mathcal{B}(0)\big|k_F^2\big)V_{\vec{k}\vec{k'}}f_{\beta,\vec{k}}(\vec{k'}),
\end{equation}
in agreement with our previous result. the Berry curvature reduces the attractive electron-electron interaction due to the Darwin term.

We are now able to calculate the zero-temperature superconducting gap. At $T=0$, the gap equation is
\begin{equation}
    \Delta_{\vec{k}}=-\frac{1}{2}\Big(1-\big|\mathcal{B}(0)\big|k_F^2\Big)\sum_{\vec{k'}}V_{\vec{k}\vec{k'}}\frac{\Delta_{\vec{k'}}}{\sqrt{\Delta_{\vec{k'}}^2+\xi_{\vec{k'}}^2}}
\end{equation}
We then use $V_{\vec{k}\vec{k'}}=-V\mathbbm{1}_{\mathcal{D}}(\vec{k})\mathbbm{1}_{\mathcal{D}}(\vec{k'})$ and have
\begin{equation}
    \Delta_{\vec{k}}=\mathbbm{1}_{\mathcal{D}}(\vec{k})\frac{1}{2}\Big(1-\big|\mathcal{B}(0)\big|k_F^2\Big)V\sum_{\vec{k}\in\mathcal{D}}\frac{\Delta_{\vec{k'}}}{\sqrt{\Delta_{\vec{k'}}^2+\xi_{\vec{k'}}^2}}
\end{equation}
Thus $\Delta_{\vec{k}}=0$ for $\vec{k}\notin\mathcal{D}$, and then one can show directly that $\Delta_{\vec{k}}=\Delta$ for $\vec{k}\in\mathcal{D}$. The former case is trivially satisfied since if $\vec{k}\notin\mathcal{D}$, the corresponding electron is not subject to the attractive interaction so it cannot condense and participate in a SC state. The latter indicates that the gap is then isotropic  for the electrons that are concerned by superconductivity. We may again follow the conventional derivation \cite{tinkham2004introduction} and find the $T=0$ superconducting gap
\begin{equation}
    \label{gaplambdaeff}
    \Delta(T=0)=\frac{\hbar\omega_D}{\sinh\big(1/\lambda_{\text{eff}}\big)}\quad\text{with}\quad\lambda_{\text{eff}}=\big(1-\big|\mathcal{B}(0)\big|k_F^2\big)\lambda
\end{equation}
with the same effective coupling constant $\lambda_{\text{eff}}$ as that obtained in the previous section [see Eq. (\ref{eq:binding})]. Comparing this to the bare BCS expression
\begin{equation}
    \Delta^{\text{BCS}}(T=0)=\frac{\hbar\omega_D}{\sinh\big(1/\lambda\big)}
\end{equation}
we see the same result as in the Cooper pair problem, that is to say a lowering of the BCS coupling constant driven by the Berry curvature thereby lowering the $T=0$ superconducting gap. This is also consistent with what we said about the consequences for the Cooper pairs. Indeed, since the superconducting gap is smaller, so is the energy of the quasiparticles in the superconductor. This makes them more sensitive to variations of energy, e.g. thermal fluctuations. In other words, the superconducting phase is weakened and thus more easily suppressed upon raising temperature.

Similarly, the expression for the critical temperature takes the form \cite{tinkham2004introduction}
\begin{equation}\label{eq:TC}
    T_c=2\hbar\omega_D\frac{e^\gamma}{\pi}e^{-1/\lambda_{\text{eff}}}
\end{equation}
and is identical to the standard one except for the fact that the coupling constant needs to be replaced by $\lambda\rightarrow \lambda_{\text{eff}}$ to take into account the extra terms due to the Berry curvature. Here, $\gamma\simeq0.577$ is the Euler-Mascheroni constant, and the approximation is valid if $2T_c\ll\hbar\omega_D/k_B=T_D$, and it is relatively reliable when $2T_c\lesssim T_D$. Notice finally, that the Berry curvature therefore does not affect the universality of the ratio between the superconducting gap and $T_c$ in the weak-coupling limit,
\begin{equation}
    \frac{\Delta(T=0)}{k_BT_c}\underset{\lambda\ll1}{=}\frac{\pi}{e^\gamma}\simeq1.76.
\end{equation}
Indeed this ratio is independent of the (effective) coupling constant.

\section{Doping dependence}\label{sec:doping}

\begin{figure}[h!]
    \centering
    \includegraphics[width=0.36\textwidth]{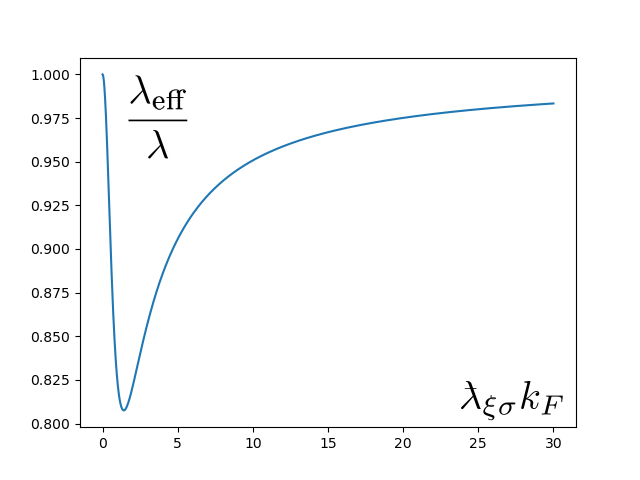}
    \caption{Ratio $\frac{\lambda_{\text{eff}}}{\lambda}$ as a function of $\lambdabar_{\xi\sigma}k_F$.}
    \label{dopingdependence}
\end{figure}

Until now, we considered a low-doping limit, in which the Fermi level is close to the bottom of the conduction band. This allowed us to approximate the Berry curvature as $\mathcal{B}(k)\simeq\mathcal{B}(0)$. At larger doping, we first expect a weakening of the inter-band effects since the relevant physics will take place farther away from the other band. We should then expect to recover the usual one-band BCS results as the Fermi energy increases. The main thing to change would be our extra terms. The Berry dipole term does not rely on the low-energy expansion of the Dirac Hamiltonian, and we thus do not need to change it. The Darwin term is different: we have obtained it by expanding the Dirac Hamiltonian in the low-energy/non-relativistic limit. In this limit, the Berry curvature enters as $\big|\mathcal{B}(0)\big|$. Since the physics is controlled by states near the Fermi energy, we change $\big|\mathcal{B}(0)\big|\longrightarrow\big|\mathcal{B}(k_F)\big|$, \textit{i.e.} the most important contribution of the Berry curvature is its value at the Fermi level. The effective coupling constant $\lambda_{\text{eff}}$ takes then the form
\begin{equation}
    \label{deforlambda}
    \lambda_{\text{eff}}=\big(1-\big|\mathcal{B}(k_F)\big|k_F^2\big)\lambda=\Bigg(1-\frac{\lambdabar_{\xi\sigma}^2k_F^2}{2\big(1+\lambdabar_{\xi\sigma}^2k_F^2\big)^{3/2}}\Bigg)\lambda,
\end{equation}
and we have 
\begin{align}
    \label{lowdopinglimit}
    &\text{Low-doping limit: }\frac{\lambda_{\text{eff}}}{\lambda}\underset{\lambdabar_{\xi\sigma}k_F\ll1}{\sim}1-\frac{\lambdabar_{\xi\sigma}^2k_F^2}{2} \\
    \label{highdopinglimit}
    &\text{High-doping limit: }\frac{\lambda_{\text{eff}}}{\lambda}\underset{\lambdabar_{\xi\sigma}k_F\gg1}{\sim}1-\frac{1}{2\lambdabar_{\xi\sigma}k_F}
\end{align}
for the different limiting cases. As a consistency check, we recover the previous result in the low-doping limit (indeed, $\lambdabar_{\xi\sigma}^2/2=\big|\mathcal{B}(0)\big|$). In the high-doping limit, the effective coupling constant approaches its bare BCS value as the Fermi level goes to $+\infty$. This is consistent with our expectation of a decreased role of the corrective terms due to the Berry curvature and thus of the inter-band effects in this limit. The doping dependence of the coupling constant (i.e. on $\lambdabar_{\xi\sigma}k_F$) is depicted in Fig. \ref{dopingdependence}.
\newline
It is apparent that the effective coupling constant has a minimum that can be shown to occur at $\lambdabar_{\xi\sigma}k_F=\sqrt{2}$. Therefore the effect of the Berry curvature on conventional BCS type ($s$-wave) superconductivity is expected to be strongest in an intermediate doping regime in which the Fermi wave vector is on the order of the inverse effective Compton length. We then have
\begin{equation}
    \min_{\lambdabar_{\xi\sigma}k_F}\frac{\lambda_{\text{eff}}}{\lambda}=1-\frac{1}{3\sqrt{3}}\simeq81\%,
\end{equation}
\textit{i.e.} the maximal reduction is approximately $19\%$. It is interesting to note that while the ratio goes to 1 as the Fermi level goes to $+\infty$, the difference does not go to zero. Indeed,
\begin{equation}
    \underset{k_F\rightarrow+\infty}{\lim}\big[\lambda_{\text{eff}}-\lambda\big]=-\frac{AV}{4\pi\Delta_b}
\end{equation}
with $A$ the area of the Brillouin zone. Note that $V$ represents, here, the interaction energy per unit area in reciprocal space so that the quantity $AV$ itself is an energy and the coupling constant is dimensionless. While the reduction of the coupling constant seems rather limited, we must not forget that the critical temperature and the superconducting gap both depend exponentially on this coupling constant, so the effect could be quite substantial. 

The central result of this paper is Eq. (\ref{deforlambda}). Indeed, from it ensues most of the results we had so far. Moreover, it could have several uses. First, doping could offer a way to experimentally observe the effects of a Berry curvature on a superconducting phase discussed in this paper. We present some possible paths for an experimental test of Berry-curvature effects on BCS superconductivity in Sec. \ref{sec:exp}. Second, while this specific deformation of the coupling constant may not be true for other types of band structures, these could still exhibit other types of deformations depending on the corrective terms of the one-body problem. If Eq. (\ref{deforlambda}) is true in other types of band structures, it can even be a way to detect the presence of a Berry curvature as well as its $k$-dependence.  

\section{Beyond BCS superconductivity}\label{sec:beyond}

Now that we have studied the conventional $s$-wave case, let us see what happens with other types of superconductivity. As in the case for the $s$-wave case (see Sec. \ref{ssec:Cooper}), we first revisit the modified Cooper problem from a more general point of view following Ref. \cite{mineev_samokhin_1999}. We will then study the many-body BCS theory, this time following Refs.  \cite{SigristUeda} and \cite{sigrist_2005}.

\subsection{Cooper problem}\label{ssec:Coopergen}

The 2-electron potential may be decomposed in the relative-angular momentum basis as \cite{mineev_samokhin_1999}
\begin{equation}
    V_{\vec{k}\vec{k'}}=\sum_{l=0}^{+\infty}V_l(\vec{k},\vec{k'})
\end{equation}
with  $V_l(-\vec{k},\vec{k'})=(-1)^lV_l(\vec{k},\vec{k'})=V_l(\vec{k},-\vec{k'})$, and the integer $l$ the angular momentum of the superconducting phase. It is even for singlet pairing and odd for triplet pairing. Let us pick a superconducting phase with fixed $l$, so that the pairing is either singlet or triplet. Then $V_{\vec{k}\vec{k'}}=V_l(\vec{k},\vec{k'})$. The same is done to $g_{\vec{k}}=g_l(\vec{k})$ with also $g_l(-\vec{k})=(-1)^lg_l(\vec{k})$. Equation (\ref{Cooper1}) becomes therefore
\begin{equation}
    \sum_{\vec{k}}\frac{\langle V^{\text{eff}}_l(\vec{k},\vec{k'})\rangle_l}{E-2\epsilon_+(\vec{k})}=1,
\end{equation}
and we then proceed in the same way as before, expanding $\langle V^{\text{eff}}_l(\vec{k},\vec{k'})\rangle_l$ and considering
\begin{equation}
    \langle\vec{k'}V_l(\vec{k},\vec{k'})\rangle_l\propto\sum_{\vec{k'}}\vec{k'}V_l(\vec{k},\vec{k'})g_l(\vec{k'}).
\end{equation}
If we then take $V_l(\vec{k},\vec{k'})$ to be non-zero only within a thin layer of energy around the Fermi level, with the energy cut-off $\epsilon_l$, and one retrieves Eq. (\ref{Dset}) but with $\mathcal{D}_l=\Big\{\vec{k}\in\mathbb{R}^2\big|\epsilon_F\leq\epsilon_+(\vec{k})\leq\epsilon_F+\epsilon_l\Big\}$. Because of the symmetry
\begin{align}
    V_l(\vec{k},-\vec{k'})g_l(-\vec{k'})&=(-1)^lV_l(\vec{k},\vec{k'})(-1)^lg_l(\vec{k'})\nonumber\\
    &=V_l(\vec{k},\vec{k'})g_l(\vec{k'}),
\end{align}
the function $\vec{k'}V_l(\vec{k},\vec{k'})g_l(\vec{k'})$ is odd in $\vec{k'}$ so that the sum over the set $\mathcal{D}_l$ yields zero. Since this term carries the Berry dipole term, we can conclude that \emph{the Berry dipole term does not contribute to the energy of the Cooper pair with pure singlet or triplet pairings}. Notice, however, that the Berry dipole term may nevertheless play a significant role in exotic superconductors that mix singlet and triplet pairing, as we sketch out in Sec. \ref{ssec:mixed}.

We then proceed with the same approximation as for the conventional $s$-wave case, which gives the effective interaction
\begin{equation}
    \langle V^{\text{eff}}_l(\vec{k},\vec{k'})\rangle_l=\big(1-\big|\mathcal{B}(k_F)\big|k_F^2\big)\langle V_l(\vec{k},\vec{k'})\rangle_l.
\end{equation}
We also take the approach of \cite{mineev_samokhin_1999} and take $V_l(\vec{k},\vec{k'})=V_l(k,k')f(\hat{k},\hat{k'})$ with $V_l(k,k')=-V_l\mathbbm{1}_{\mathcal{D}_l}(\vec{k})\mathbbm{1}_{\mathcal{D}_l}(\vec{k'})$. This approach gives a binding energy $E_{B,l}$ given by
\begin{equation}
    E_{B,l}=\frac{2\epsilon_l}{e^{2/\lambda_{\text{eff}}}-1}
\end{equation}
with $\lambda_{\text{eff}}=\big(1-\big|\mathcal{B}(k_F)\big|k_F^2\big)\lambda$, i.e. \emph{the result obtained for the conventional Cooper problem extends to all singlet and triplet pairings}.

\subsection{Many-body problem: generalized BCS theory}\label{ssec:BCSgen}

We now briefly address the many-body problem from a more general point of view, using the generalized BCS theory presented in Refs. \cite{sigrist_2005} and \cite{SigristUeda}. Its Hamiltonian is
\begin{equation}
    H=\sum_{\vec{k}}\xi_{\vec{k}}c^\dagger_{\vec{k}\sigma}c_{\vec{k}\sigma}+\frac{1}{2}\sum_{\substack{\sigma_1\sigma_2\\\sigma_3\sigma_4}}\sum_{\vec{k}\vec{k'}}V^{\substack{\sigma_1\sigma_2\\\sigma_3\sigma_4}}_{\text{eff},\vec{k}\vec{k'}}c^\dagger_{\vec{k}\sigma_1}c^\dagger_{-\vec{k}\sigma_2}c_{-\vec{k'}\sigma_3}c_{\vec{k'}\sigma_4},
\end{equation}
with the effective interaction containing the Berry curvature corrections. The mean-field theory of this Hamiltonian gives rise to a $2\times2$ matrix $\widehat{\Delta}_{\vec{k}}$. As in the conventional case, one can prove that the dressed order parameter has the same parity as the bare one. Similarly to the Cooper problem, let us investigate a pairing that is either singlet or triplet. Then the gap equation has the form \cite{SigristUeda} 
\begin{equation}
    \Delta^{\sigma_1\sigma_2}_{\vec{k}}=-\sum_{\sigma_3\sigma_4}\sum_{\vec{k'}}V^{\substack{\sigma_2\sigma_1\\\sigma_3\sigma_4}}_{\text{eff},\vec{k}\vec{k'}}\mathscr{I}^{\sigma_3\sigma_4}_\beta(\vec{k'}),
\end{equation}
and the expansion of the effective interaction yields 
\begin{widetext}
\begin{align}
    \label{expandVgen}
    &\Delta^{\sigma_1\sigma_2}_{\vec{k}}=-\bigg(1-\frac{1}{2}\big|\mathcal{B}(k_F)\big|\vec{k}^2\bigg)\sum_{\sigma_3\sigma_4}\sum_{\vec{k'}}V^{\substack{\sigma_2\sigma_1\\\sigma_3\sigma_4}}_{\vec{k}\vec{k'}}\mathscr{I}^{\sigma_3\sigma_4}_{\beta}(\vec{k'})-\Bigg(\frac{i}{2}\vec{\Lambda}^{\xi_1,\xi_2}_-(\vec{k})\times\vec{k}+\big|\mathcal{B}(k_F)\big|\vec{k}\Bigg)\cdot\sum_{\sigma_3\sigma_4}\sum_{\vec{k'}}\vec{k'}V^{\substack{\sigma_2\sigma_1\\\sigma_3\sigma_4}}_{\vec{k}\vec{k'}}\mathscr{I}^{\sigma_3\sigma_4}_{\beta}(\vec{k'})\nonumber\\
    &-\frac{1}{2}\big|\mathcal{B}(k_F)\big|\sum_{\sigma_3\sigma_4}\sum_{\vec{k'}}\vec{k'}^2V^{\substack{\sigma_2\sigma_1\\\sigma_3\sigma_4}}_{\vec{k}\vec{k'}}\mathscr{I}^{\sigma_3\sigma_4}_{\beta}(\vec{k'}),
\end{align}
\end{widetext}
where the summand of the $\vec{k'}$-linear term is \begin{equation}
    \sum_{\sigma_3\sigma_4}\vec{k'}V^{\substack{\sigma_2\sigma_1\\\sigma_3\sigma_4}}_{\vec{k}\vec{k'}}\mathscr{I}^{\sigma_3\sigma_4}_\beta(\vec{k'}).
\end{equation}

We study two separate cases now. First, let us consider a unitary pairing, \textit{i.e.} one for which $\widehat{\Delta}_{\vec{k}}\widehat{\Delta}^\dagger_{\vec{k}}\propto\sigma_0$. This entails all singlet pairings and unitary triplet pairings (those without spin polarization). In that case, the kernel $\hat{\mathscr{I}}_{\beta}(\vec{k'})$ is given by \cite{SigristUeda,sigrist_2005}
\begin{equation}
    \hat{\mathscr{I}}_\beta(\vec{k'})=\frac{\widehat{\Delta}_{\vec{k'}}}{2E_{\vec{k'}}}\tanh\bigg(\frac{\beta}{2}E_{\vec{k'}}\bigg).
\end{equation}
Since the order parameter generally obeys $\widehat{\Delta}_{-\vec{k}}=-\widehat{\Delta}^\top_{\vec{k}}$ and $E_{-\vec{k}}=E_{\vec{k}}$, we have 
\begin{equation}
    \mathscr{I}^{\sigma_3\sigma_4}_\beta({-\vec{k'}})=-\mathscr{I}^{\sigma_4\sigma_3}_\beta(\vec{k'}).
\end{equation}
Furthermore, in order to respect the anticommutation relations of the fermionic operators, the interaction must obey $V^{\substack{\sigma_2\sigma_1\\\sigma_3\sigma_4}}_{\vec{k};-\vec{k'}}=-V^{\substack{\sigma_2\sigma_1\\\sigma_4\sigma_3}}_{\vec{k}\vec{k'}}$ \cite{sigrist_2005}. With this, we have
\begin{align}
    \sum_{\sigma_3\sigma_4}-\vec{k'}V^{\substack{\sigma_2\sigma_1\\\sigma_3\sigma_4}}_{\vec{k};-\vec{k'}}\mathscr{I}^{\sigma_3\sigma_4}_\beta(-\vec{k'})&=-\sum_{\sigma_3\sigma_4}\vec{k'}V^{\substack{\sigma_2\sigma_1\\\sigma_4\sigma_3}}_{\vec{k}\vec{k'}}\mathscr{I}^{\sigma_4\sigma_3}_\beta(\vec{k'})\nonumber\\
    &=-\sum_{\sigma_3\sigma_4}\vec{k'}V^{\substack{\sigma_2\sigma_1\\\sigma_3\sigma_4}}_{\vec{k}\vec{k'}}\mathscr{I}^{\sigma_3\sigma_4}_\beta(\vec{k'})
\end{align}
\textit{i.e.} the latter is odd in $\vec{k'}$. If one takes the interaction to be non-zero in a thin layer of energy around the Fermi level with energy cutoff $\epsilon_c$, the sum over the term that is linear in $\vec{k'}$ in Eq. (\ref{expandVgen}) vanishes again, a situation encountered several times in this paper. So the Berry dipole term does not change the gap and critical temperature for unitary pairings. As pointed out in the appendix, the latter is also valid for non-unitary triplet pairings. Therefore, \emph{the Berry dipole term does not change the gap equation for pure singlet and triplet pairings}.

\subsection{Possible situations in which the Berry dipole term may become relevant}\label{ssec:mixed}

In view of the above results, one may then wonder if there is any possible effect of the Berry dipole term on superconductivity. What we proved so far is that it does not change the SC gap or $T_c$ if the parity of the pairing is well defined. So a necessary condition for the Berry dipole term to actually contribute would be a superconducting phase without a fixed parity. We saw in the Cooper problem that the Berry dipole term drops out because the following sum is zero
\begin{equation}
    \label{sumVg}
    \sum_{\vec{k'}}\vec{k'}V_{\vec{k}\vec{k'}}g_{\vec{k'}}.
\end{equation}
If we decompose the two functions $V_{\vec{k}\vec{k'}}$ and $g_{\vec{k'}}$ in the sum of an even and an odd function
\begin{align}
    V_{\vec{k}\vec{k'}}&=V^e_{\vec{k}\vec{k'}}+V^o_{\vec{k}\vec{k'}}\\
    g_{\vec{k'}}&=g^e_{\vec{k'}}+g^o_{\vec{k'}}
\end{align}
and interpret the $e$ and $o$ parts respectively as the singlet and triplet parts, we then have
\begin{equation}
    V_{\vec{k}\vec{k'}}g_{\vec{k'}}=V^e_{\vec{k}\vec{k'}}g^e_{\vec{k'}}+V^o_{\vec{k}\vec{k'}}g^o_{\vec{k'}}+V^o_{\vec{k}\vec{k'}}g^e_{\vec{k'}}+V^e_{\vec{k}\vec{k'}}g^o_{\vec{k'}}.
\end{equation}
While the first two terms disappear in Eq. (\ref{sumVg}) as they are even functions of $\vec{k'}$, the other two terms do \textit{a priori} not disappear as they are odd functions of $\vec{k'}$. $V^og^e$ may be interpreted as the interactions between triplet pairs in the presence of singlet pairs while $V^eg^o$ is the opposite. These two terms may then be an opportunity for the Berry dipole term to have a non-zero contribution in the superconducting phase, \textit{i.e.} if the latter shows coexistence between singlet and triplet pairs. We would then need a superconducting phase where none of the two dominates. Some materials have been proposed to exhibit two superconducting phases, each with a different parity, such as $\text{CeRh}_2\text{As}_2$ and bilayer-$\text{NbSe}_2$  \cite{Mockli2022, Khim2021}. Notice furthermore that a very recent theoretical study argues that the observed superconducting phase in twisted bilayer graphene \cite{Cao2018} might be due to an admixture of singlet and triplet pairs \cite{Lake2022}, and the Berry dipole term might then be a relevant parameter in the stabilization of this type of superconductivity. Note that for both $\text{CeRh}_2\text{As}_2$ and bilayer-$\text{NbSe}_2$, and generally in non-centrosymmetric superconductors \cite{Frigeri2006}, a magnetic field is necessary to obtain a parity-mixed superconducting phase. To circumvent this issue, one could first implement the magnetic field in a BCS formalism for these systems, thereby absorbing it in an effective superconducting order parameter/interaction, which could then be used to our discussion here. This approach seems plausible since the coupling between the Berry curvature and the magnetic field appears in the equations of motion as the product of the two, so as a second term term that could be neglected as long as the magnetic field necessary is not too high.

\section{Possible experimental implications of the Berry curvature on 2D BCS superconductivity}\label{sec:exp}

\begin{figure}[htbp]
    \centering
    \includegraphics[width=\linewidth]{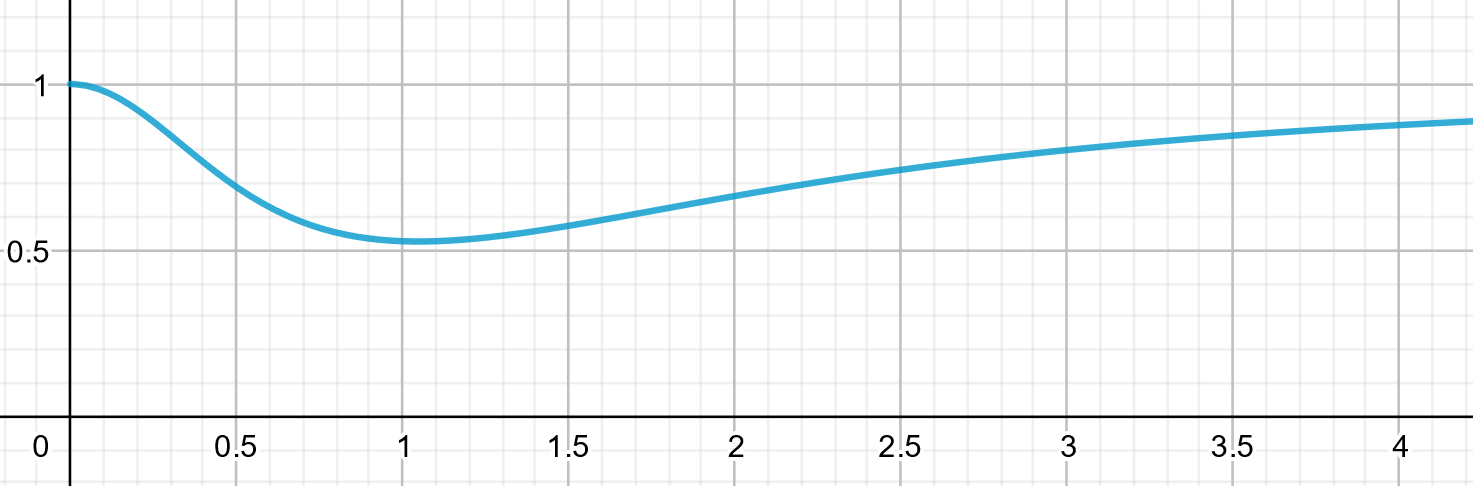}
    \caption{Ratio $T_c/T_c^{\text{BCS}}$ as a function of $\lambdabar_{\xi\sigma}k_F\propto \sqrt{n_{\text{2D}}}$. Here, we have used $AV/2\pi\Delta_{\xi\sigma}\simeq0.2$ for illustration.}.
    \label{fig:my_label}
\end{figure}

As shown in Sec. \ref{sec:doping}, the Berry curvature has its strongest effect at Fermi wave vectors that are on the order of the inverse effective (Compton) length $\lambdabar_{\xi\sigma}$. Even if the relative reduction of the coupling constant is on the order of 19\%, one needs to keep in mind that the experimentally measurable superconducting gap and critical temperature depend exponentially on the coupling constant. Indeed, the former is accessible by spectroscopic means, \textit{e.g.} in scanning-tunneling spectroscopy, and the latter within resistive temperature-dependent measurements. Experimentally, it is likely impossible to change the Berry curvature \textit{in situ} because this would require experimental access to the band parameters, such as the direct band gap in 2D TMDC. While one could hope to change it \textit{e.g} under strain, also the phonon spectrum and the electron-phonon coupling would then change, possibly in an uncontrolled manner, thus excluding a direct measurement of the Berry-curvature effect in superconductivity. 

However, one may compare the evolution of the Berry-curvature dependent superconducting gap or critical temperature, measured as a function of doping, to the \textit{expected} behavior of these quantities. Direct comparison of the critical temperature $T_c$ in Eq. (\ref{eq:TC}), in terms of the effective coupling constant (\ref{eq:binding}), yields the ratio
\begin{equation}\label{eq:ratioTC}
    \frac{T_c}{T_c^{\text{BCS}}}=\exp\Bigg(-\frac{2\pi\Delta_{\xi\sigma}}{AV}\frac{\big|\mathcal{B}(k_F)\big|k_F^2}{\Big(1-\big|\mathcal{B}(k_F)\big|k_F^2\Big)\sqrt{1+\lambdabar_{\xi\sigma}^2k_F^2}}\Bigg),
\end{equation}
where $T_c^{\text{BCS}}$ is the BCS critical temperature in the absence of Berry-curvature terms. 
We notice here the clear competition between the Berry curvature (through the gap) and superconductivity (through the attractive interaction $V$.). The ratio (\ref{eq:ratioTC}) is plotted in Fig. \ref{fig:my_label} as a function of the doping-dependent Fermi wave vector, $k_F=\sqrt{(4\pi/g) n_{\text{2D}}}$, in terms of the induced 2D electronic density $n_{\text{2D}}$. The factor $g$ takes into account the degeneracy due to internal degrees of freedom, such as the valley and the spin. Notice that, in 2D TMDC with a prominent spin-orbit coupling, the valley and spin degrees of freedom are generically locked, as mentioned above. One would therefore expect $g=2$ in these materials. This is likely the case in the valence band, with a spin-orbit splitting on the order of $\sim 100$ meV, while it is only in the $\sim 1...10$ meV range in the conduction band. The reduction of the critical temperature is strongest at the minimum, which occurs at $\lambdabar_{\xi\sigma}k_F\simeq1.05$. This corresponds to an electronic density of
\begin{equation}
    n_{2\text{D}}=\frac{g}{4\pi}k_F^2\simeq1.1\frac{g}{4\pi}\lambdabar_{\xi\sigma}^{-2}.
\end{equation}
We can then give an approximation of the minimum of the ratio as
\begin{equation}
    \min_{k_F}\frac{T_c}{T_c^{\text{BCS}}}\simeq\exp\bigg(-0.15\frac{2\pi\Delta_{\xi\sigma}}{AV}\bigg).
\end{equation}

\section{Conclusions}

In conclusion, we have studied the effect of the Berry curvature on BCS-type superconductors in 2D electronic systems. We have shown that the two-body Hamiltonian for interacting electrons inherits terms that are linear in the Berry curvature and that are inherited from the single-electron band structure. In this case, the Berry curvature, which arises in the adiabatic limit when the electrons are restricted to a single band due to purely virtual transitions to the other bands, is coupled to electric potentials beyond the periodic one, which gives rise to the Bloch bands. While such potentials may arise due to external electric fields, they naturally arise when interactions between the electrons (or holes) are taken into account. Generically, the Berry curvature provides a dipolar structure to the charged pairs, and one of the terms emerging in the two-body Hamiltonian can indeed be interpreted as a dipole in an electric field. A second term emerges in the form of a Darwin term, in which the Berry curvature couples to the Laplacian of the electric potential. This term is best understood within a relativistic treatment of the (massive) Dirac Hamiltonian that mimics the two adjacent bands in a direct-gap semiconductor. 

Following the lines of the usual BCS treatment of superconductivity in the weak-coupling limit, we have shown that the latter Darwin term generally lowers the BCS coupling constant. As a consequence, this lowers also the stability of the Cooper pair so that the superconducting gap and critical temperature are decreased. On the contrary, the dipolar term, which potentially has the power to increase superconductivity, does not affect the superconducting properties in an $s$-wave or any pure singlet or triplet superconductor because of their fixed parity. The dipolar term might then play a role in systems where superconducting phases of different parity coexist or where the superconducting order parameter does not have a fixed parity. This path might be explored in future work, but it is beyond the scope of our present paper. 

Interestingly, the gap-to-$T_c$ ratio remains the same as in the conventional BCS theory in the weak-coupling limit, that we have considered here. Upon doping, the reduction of BCS superconductivity is strongest when the Fermi wave vector is on the order of the inverse effective Compton length, $k_F\sim\lambdabar_{\xi\sigma}^{-1}$,  where the BCS coupling constant is lowered by $19\%$. Indeed, for stronger doping, the Fermi level is situated at wave vectors, where the Berry curvature rapidly tends to zero. Since the superconducting gap and the critical temperature both depend exponentially on the BCS coupling constant, the relatively weak reduction of the coupling constant is more prominent there. Our calculations show that the reduction of the doping-dependent superconducting gap and critical temperature depends then both on the band gap, which determines the value of the Berry curvature, as well as on the effective electron-electron interaction. The experimental measurement of these quantities in 2D materials upon doping might then provide a test of our theoretical studies if compared to the expected evolution predicted by the usual BCS theory in the absence of Berry-curvature corrections.

\section*{Acknowledgements}
We thank J. Meyer and A. Mesaros for valuable discussions.

\bibliographystyle{apsrev4-2}
\bibliography{Biblioarticle}


\appendix\label{sec:appA}
\begin{widetext}
\section{Gap equation for non-unitary pairings}
We take the gap equation 
\begin{equation}
    \Delta^{\sigma_1\sigma_2}_{\vec{k}}=-\sum_{\sigma_3\sigma_4}\sum_{\vec{k'}}V^{\substack{\sigma_2\sigma_1\\\sigma_3\sigma_4}}_{\text{eff},\vec{k}\vec{k'}}\mathscr{I}^{\sigma_3\sigma_4}_\beta(\vec{k'}).
\end{equation}
and pick a non-unitary triplet pairing. In that case, the kernel is given by \cite{SigristUeda}
\begin{equation}
    \hat{\mathscr{I}}_\beta(\vec{k})=i\vec{\alpha}_\beta(\vec{k})\cdot\vec{\sigma}\sigma_y
\end{equation}
with
\begin{equation}
    \vec{\alpha}_\beta(\vec{k})=\frac{1}{2E_{\vec{k},+}}\bigg(\vec{d}(\vec{k})+\frac{1}{|\vec{q}(\vec{k})|}\vec{d}(\vec{k})\times\vec{q}(\vec{k})\bigg)\tanh\bigg(\frac{\beta}{2}E_{\vec{k},+}\bigg)+\frac{1}{2E_{\vec{k},-}}\bigg(\vec{d}(\vec{k})-\frac{1}{|\vec{q}(\vec{k})|}\vec{d}(\vec{k})\times\vec{q}(\vec{k})\bigg)\tanh\bigg(\frac{\beta}{2}E_{\vec{k},-}\bigg).
\end{equation}
Moreover, $\vec{q}=i\vec{d}\times\vec{d}^*$ and $E_{\vec{k},\pm}=\sqrt{\xi_{\vec{k}}^2+|\vec{d}(\vec{k})|^2\pm|\vec{q}(\vec{k})|}$. Since $\vec{d}$ is an odd function of $\vec{k}$ and $E_{\vec{k},\pm}$ an even one, $\vec{\alpha}_\beta$ is an odd function of $\vec{k}$ and therefore the kernel $\hat{\mathscr{I}}$ as well. Another thing to notice is that expliciting the matrix form of the kernel yields
\begin{equation}
    \hat{\mathscr{I}}_\beta(\vec{k})=\begin{pmatrix}
    -\alpha_{\beta,x}(\vec{k})+i\alpha_{\beta,y}(\vec{k})&\alpha_{\beta,z}(\vec{k})\\
    \alpha_{\beta,z}(\vec{k})&\alpha_{\beta,x}(\vec{k})+i\alpha_{\beta,y}(\vec{k})
    \end{pmatrix}
\end{equation}
which is obviously a symmetric matrix. Putting the two together, we have
\begin{equation}
    \mathscr{I}^{\sigma_3\sigma_4}_\beta(-\vec{k})=-\mathscr{I}^{\sigma_3\sigma_4}_\beta(\vec{k})=\mathscr{I}^{\sigma_4\sigma_3}_\beta(\vec{k})
\end{equation}
which is what we used to show that the Berry dipole term does not change the gap or the critical temperature. So the latter also extends to non-unitary triplet pairings.
\end{widetext}
\end{document}